\PassOptionsToPackage{hyphens}{url}
\documentclass[12pt]{spieman}  
\usepackage{amsmath,amsfonts,amssymb}
\usepackage{graphicx}
\usepackage{setspace}
\usepackage{tocloft}
\usepackage{lineno}
\usepackage{hyperref}

\usepackage[style=authoryear,backend=biber]{biblatex}

\title{Causal Stance}

\author[a*]{Yoshiyuki Ohmura}
\author[a]{Yasuo Kuniyoshi}

\affil[a]{The University of Tokyo, Graduate School of Information Science and Technology  , Department of Mechano-Informatics, 7-3-1 Hongo, Bunkyo-ku, Tokyo, Japan}

\cftpagenumbersoff{figure}
\cftpagenumbersoff{table} 
\begin{document} 
\maketitle

\begin{abstract}
What is the meaning of physical causal closure? Jaegwon Kim explicitly adopts a conception of causation according to which physical causation is effectively identified with deterministic physical lawfulness, and equates it with physical determinism. 
While this conception is internally coherent, it differs from currently dominant theories of causation. 
Physics and the theory of causation serve different descriptive purposes. In this study, we refer to them, respectively, as the Physical Stance and the Causal Stance. Within this framework, physical determinism belongs to the Physical Stance, and physical causal closure is defined only within the Causal Stance. Consequently, the two should not be equated. 
On this basis, this study reconstructs Davidson’s anomalous monism as a materialist position that acknowledges mental causation without contradicting physical determinism. Furthermore, we propose a linguistic framework in which physical causal closure does not hold in the Causal Stance while physical determinism remains intact in the Physical Stance.
\end{abstract}


{\noindent \footnotesize\textbf{*}Yoshiyuki Ohmura,  \linkable{ohmura@isi.imi.i.u-tokyo.ac.jp} }

\begin{spacing}{1}   

\section{Introduction}
Materialism assumes that the mind is physical. 
On the other hand, the mind-body problem posits that the mind is not physical.
To resolve this contradiction, we should consider that ``physical'' has different meanings depending on the context.
We propose changing the meaning of ``physical'' in physics and the theory of causality.

In the philosophy of science, there is a view that science models reality using different languages for each distinct field (\cite{fraassen, giere}). These different languages are employed to address the unique challenges of each academic discipline. However, in emerging fields where a common linguistic framework has not yet been established, there is a risk of misunderstanding. This is particularly true in disciplines dealing with the mind and consciousness, where the lack of a common language can make it difficult to accurately comprehend others' arguments.

If we regard physics as the most fundamental discipline and view psychology and sociology as applied sciences, there is no major problem with interpreting the mind and mental causation in terms of folk psychology within these applied fields. 
The reason is that the purpose of these disciplines does not necessarily involve questioning the relationship between the mind and the physical world. 
Furthermore, as long as it aligns with the discipline’s purpose, the inclusion of the mind or mental causation within the language used by the discipline does not lead to confusion. 
However, in fields such as the science of consciousness, which examine the relationship between the physical and the mind, the meaning of the concepts of the mind and mental causation becomes an issue. 
There is no common understanding of these concepts, including whether they should be included in the language of consciousness studies. 
Nevertheless, opinions on whether the mind and mental causation should be introduced into the discipline’s common language will likely vary depending on whether the science of consciousness is viewed as a branch of physics or as a distinct academic field. 
Naturally, if one views it as a branch of physics, the view would likely be that the mind should not be included in that language. 
It is important to note here that naturalism and physicalism do not necessarily hold that the science of consciousness is a branch of physics.

This study argues that the science of consciousness is difficult to explain using the language of physics alone, and that it is necessary to introduce, at the very least, the language of the theory of causation. 
Following \cite{dennett}, this study refers to the interpretation using the language of physics as the ``Physical Stance,'' and the interpretation using the language of causation as the ``Causal Stance.'' 
It is well known that physics does not distinguish cause and effect as asymmetrical roles (\cite{blanchard}), and it is currently believed that the theory of causation must extend the language of physics (\cite{pearl}). 
This study aims to highlight that the Physical Stance and the Causal Stance have not been clearly distinguished in discussions concerning mental causation in the philosophy of mind.

In the philosophy of mind, Kim's exclusion  argument (\cite{kim1998}) is frequently discussed (\cite{gebharter,hoffmann, kinney,  stern, steward2012, woodward2015}). However, these discussions pay little attention to the fact that Kim equates physical causal closure with 
physical determinism. 
Kim explicitly adopts a conception of causation according to which physical causation is effectively identified with deterministic physical lawfulness. On this conception, the claim of physical causal closure can be understood as the claim that the physical state of the world at any time is fully determined by prior physical states in accordance with physical laws.
However, while the currently dominant view of causality based on manipulability emphasizes the asymmetry between cause and effect, since such asymmetry is not included in the descriptions of physics, we believe we should distinguish between physical determinism and physical causal closure.

While the physical determinism can be understood within the framework of the Physical Stance, the physical causal closure cannot. 
This has rarely been emphasized in past discussions, but physics lacks the language necessary to define physical causal closure. 
The term ``physical causal closure'' can only be defined within the framework of the Causal Stance.
Physics requires that physical determinism hold, rather than physical causal closure.
We should not equate physical determinism with physical causal closure.

Just as ``causality'' is an undefined term in physics, different academic disciplines may have different linguistic frameworks. 
Concepts that cannot be distinguished in physics may have meaning within different linguistic frameworks. 
Furthermore, the same word may have different meanings in different linguistic frameworks. 

If physics does not rule out mental causation on the basis that causation is undefined, there should be scope in other disciplines to develop models that allow for it. 
In this study, we define the meanings of ``Physical Laws'' in the Physical Stance as P-Physical Laws and those in the Causal Stance as C-Physical Laws. Furthermore, we introduce C-Mental Laws in the Causal Stance. In the similar way, we can define P-Physical States, C-Physical States, and C-Mental States. 
Through this classification, we explore the possibility of defining C-Mental Causation within the Causal Stance. 
We also explore the possibility of constructing a linguistic system in which C-Physical Causal Closure is false, given this classification. 
We believe  constructing such a linguistic system is necessary because it would explain how our folk psychological intuitions are formed. 
Additionally, we believe that the science of consciousness and psychology require a linguistic system in which consciousness and mental causation have meaning (\cite{kleiner2023}).

This study reconstructs Anomalous Monism (\cite{davidson1970}) by distinguishing between the Causal Stance and the Physical Stance. 
Anomalous Monism was developed to defend mental causation within the framework of monism; by clearly distinguishing between the Causal Stance and the Physical Stance, its claims become clearer and can serve as a foundation for constructing a linguistic system that accounts for mental causation.

The Anomalous Monism we are reconstructing differs from Davidson’s in the following respects. (1) To clarify the relationship between C-Mental and C-Physical, we will define their exact relationship. (2) We consider C-Mental Laws to be laws, although they are not as strict as P-Physical Laws. (3) Furthermore, by distinguishing between P-Physical Laws and C-Physical Laws, we resolve the contradiction arising from the idea that mental laws are not strict and physical laws are strict.

One might argue that the issue of description is nonessential. However, in applied science, particularly in engineering design, the direction of causality is of paramount importance and plays an essential role. 
Furthermore, in causal inference, it is extremely important to take the direction of causality into account (\cite{pearl}). 
In general, when the assumptions underlying a model change, the scientific methods used must also change.
Converting a causal model into a dynamical system model is possible by ignoring the asymmetry based on manipulability. However, converting a dynamical system model into a causal model requires identifying such asymmetry,  for which there is currently no established method.
For this reason, we believe clearly distinguishing between the Causal and Physical Stances is practical in science.

\section{Physics and Causality}
This paper does not claim that physics denies causation; rather, it claims that physics, given its descriptive and practical aims, does not explicitly distinguish causal asymmetries.

The physical laws described by physics exhibit a high degree of temporal symmetry and cannot account for asymmetries such as those between cause and effect. In recent years, thanks to the contributions of \cite{pearl} and \cite{woodward}, the differences between the language of physics and the language of causality have become clear. 
According to \cite{pearl}, it is essential to introduce an asymmetry that distinguishes between cause and effect based on manipulability in order to perform causal inference.

For example, in Newton’s laws, there is a relationship between force $f$, acceleration $a$, and mass $m$, expressed as $f = ma$. In this context, changing the force to alter the acceleration is a natural approach from the perspective of manipulability, and can be described as $a := f/m$. Here, $:=$ is an assignment operator (\cite{pearl}) introduced into causality theory to describe asymmetry; it is not necessary when describing the laws of physics. 

Unlike the equality sign $=$, this is an asymmetric relation: if $a := f/m$, it does not necessarily follow that $f/m := a$ . 
In practice, it is generally difficult to control force through acceleration. 
A thermometer’s reading changes with temperature, but the reading itself does not change the temperature from the perspective of manipulability. Although this phenomenon follows the laws of physics, physics does not always explain these asymmetries.
Introducing asymmetry in causality does not mean introducing non-physical forces; it simply means that, for the purpose of constructing physics, there is no value in introducing asymmetry based on manipulability.
Of course, it is possible that the language of physics may evolve in the future with the aim of explaining the physical principles underlying this asymmetry based on manipulability.

The important point is that just because the theory of causation cannot be described by the language of physics does not mean it introduces energies or forces that cannot be explained by physics, nor does it contradict the physical determinism. 
It is simply that different scientific disciplines possess different linguistic systems and semantics because they serve different purposes. 
Furthermore, because the theory of causation incorporates the concept of new asymmetries, 
models that are indistinguishable under the Physical Stance become distinguishable under the Causal Stance. 
In physics, even when an interaction is interpreted as occurring between A and B, 
in the theory of causation, it is possible that A is the cause and B the effect, 
or that B is the cause and A the effect, or that it is a physical interaction. 
Therefore, we believe that distinguishing between the Physical Stance and the Causal Stance when examining the problem of mental causation is valuable.

\section{Physical Causal Closure}
What exactly does ``physical causal closure'' mean? 
In the Physical Stance, the term “physical causal closure” is undefined  because causation is undefined.

\cite{kim1998} criticized mental causation by assuming physical causal closure. 
He held that mental states supervene on physical states. 
The relationship between supervenient and subvenient entities is that changes in supervenient entities cannot occur without changes in subvenient states; Kim considered this to be the most promising characterization for explaining the mind on the basis of physicalism.
He argued that if sufficient causal relations hold between physical properties and other physical properties, and if mental properties supervene on physical properties, then the mind has no causal power over the physical world. 
The problem here is that in physics, 
the relationship between physical states  is not causal in the sense of interventionist asymmetry
 (\cite{mathias}). The physical states at time $T+1$ is determined by the physical states at time $T$, and in reversible systems, the physical states at time $T$ is also determined by the physical states at time $T+1$. 

Relationships between physical states at different times obey physical laws, but these are not causal relationships in the interventionist account. 
Furthermore, the relationship between the mind and physics, or supervenience, is not causal.
And if, as Kim argues, mental causation does not exist either, then nothing related to causation remains in his argument. 
Thus, if one adopts the Physical Stance, the terms ``physical causation,'' ``mental causation,'' and ``physical causal closure'' are all undefined concepts.

Do these terms have any meaning in the Causal Stance? 
If we adopt the Causal Stance, it may be possible to define physical and mental causation, as long as it does not contradict the idea that P-Physical States obey P-Physical Laws and are solely determined by P-Physical States at different time in the Physical Stance. 
There appears to be no logical contradiction between physics excluding non-physical forces under the Physical Stance and the theory of causation distinguishing between physical and mental causation under the Causal Stance, provided that an interpretation is possible between the two.

\section{Anomalous Monism}
Anomalous Monism is a philosophical position proposed by \cite{davidson1970, davidson1990}, consisting of three assumptions: (1) mental events are causally related to physical events; (2) causal relations follow strict (physical) laws of determination; and (3) there are no strict laws that predict or explain mental events. 
Here, we define a strict law as a universal law—one that remains constant regardless of the individual. Because mental laws vary greatly from person to person, they cannot be considered strict laws. 
Examining each of these points: (1) is a description based on the Causal Stance. From this, we can say that mental events and physical events are terms that should be interpreted within the Causal Stance. Next, (2) refers to the fact that the theory of causation does not contradict physics. In other words, it is a condition regarding the consistency between the Physical Stance and the Causal Stance. 
This suggests that mental causation is a causal mechanism that conforms to the laws of physics.
Since (3) refers to mental events, we believe it should be interpreted within the Causal Stance.
The key point is that, in the Physical Stance, mental and physical events are interpretable as changes in physical states, but they can be discriminated in the Causal Stance. 
In the Causal Stance, they do not necessarily follow strict laws.

\cite{kim1998} criticized Anomalous Monism by firmly linking causation to the laws of physics. However, his conception of causation differs from the one widely accepted in science today. Davidson’s conception of causation, which makes a clear distinction between causation and laws, is considered to be more compatible with the modern view of causation.

Thus, Davidson’s Anomalous Monism, within the Causal Stance, asserts that when a causal model exists and defines mental and physical events, (1) mental events have causal power over physical events, and (3) mental events are not determined by strict laws in the Causal Stance. 
Here, to emphasize that these are physical laws satisfied by physical events within the Causal Stance, we will refer to them as ``C-Physical Laws'', and mental laws as ``C-Mental Laws''. 
In this case, since C-Physical Laws are subject to mental causation, changes in the states of C-Physical States are not determined by C-Physical States alone. 
Furthermore, the physical laws pertaining to point (2) within the Physical Stance are defined as ``P-Physical Laws''.
The aim of this study is to enable the meaning of ``Physical Laws'' to be modified according to each Stance by distinguishing between the Physical Stance and the Causal Stance.

Here, since C-Mental States supervene on C-Physical States, C-Mental States do not change without a change in C-Physical States. 
Davidson argued that the supervenience-subvenience relationship does not follow strict laws. 
However, since we view the supervenience-subvenience relationship as a mereological whole-parts relation, we consider this relationship to be fixed.
Therefore, if C-Mental States do not satisfy strict laws, C-Physical States should likewise not satisfy strict laws. 
In other words, changes in C-Physical States are not determined solely by C-Physical States but are also influenced by C-Mental States, meaning that the C-Physical Laws governing these changes are not strict laws.
Unlike Davidson, we attempt to resolve the contradiction between the fact that mental laws do not follow strict laws and physical deterministic lawfulness by distinguishing between the Physical Stance and the Causal Stance.

It is important to note here that unless it is made clear that the C-Physical Laws and the P-Physical Laws  have different meanings, our argument will contain an internal contradiction. This is because, the C-Physical Laws do not satisfy the strict laws stated in (3), but the P-Physical Laws are assumed to be determined by the strict physical laws as stated in (2). Unless we distinguish between the Physical Stance and the Causal Stance, physical laws are both strict laws and not strict laws, which is a contradiction.

Given this distinction, if a causal model that satisfies condition (2) can also satisfy conditions (1) and (3), then there is no logical contradiction in the claim of reconstructed Anomalous Monism. In the next section, we will discuss whether it is possible to construct a causal model that supports Anomalous Monism.

\section{Dual-Laws Model for Anomalous Monism}
It should be emphasized that the Dual-Laws Model is introduced here not as a unique or necessary realization of Anomalous Monism, but merely as one possible model that makes its interpretation concrete within the Causal Stance. The central claim of this paper is not tied to any particular model, but rather to the distinction between the Physical Stance and the Causal Stance, which allows Anomalous Monism to be consistently understood without generating a contradiction with physical determinism.

The conversion from the Causal Stance to the Physical Stance can be achieved by replacing the causal assignment operator $:=$ with an equals sign $=$. However, it is more difficult to identify causal asymmetry based on manipulability within the supervenience-subvenience model.

In this section, we argue that the Dual-Laws Model satisfies the conditions of Anomalous Monism, as reconstructed in the previous section. This allows us to construct a model within the Causal Stance that distinguishes between C-Physical Laws and C-Mental Laws and does not satisfy C-Causal Physical Closure. On the other hand, we argue that when interpreted within the Physical Stance, the Dual-Laws Model satisfies physical determinism, since all changes in P-Physical States are determined solely by P-Physical States.

\subsection{Dual-Laws Model (DLM)}
The Dual-Laws Model assumes a hierarchical system consisting of a layer of supervenient entities and a layer of the corresponding subvenient entities. 
Each of the supervenience layer and the subvenience layer possesses independent dynamics, and through inter-level causation, causes arising in the supervenience layer alter the states of the subvenient entities. 
In this case, causal transmission is unidirectional, and the reverse transmission does not hold. 
Therefore, when distinguishing models based on the direction of causal transmission using the semantics of causation, this causal model can be distinguished from a single-layer dynamical model in physics.

Causality is defined by the asymmetry of manipulability. Manipulating a cause affects the effect, but the reverse is not true. The DLM is a model designed to introduce this asymmetry between causes and effects across hierarchical levels.

In this model, the supervenience layer is composed of multiple supervenient functions. 
Furthermore, we define supervenient causes as changes in equations constructed by selecting and ordering subsets  of supervenient functions. 
The causal transmission mechanism that conveys the causes generated in this way to the subvenience level is a self-referential feedback mechanism. 
This mechanism observes feedback errors and modifies the states of subvenient entities to reduce those errors, thereby satisfying the specified equations. These equations serve as error functions for calculating feedback error.
For details, please refer to the formulation in the Appendix.

In the DLM, independent dynamical laws are assumed at the supervenience level and the subvenience level, respectively, but these are not causal relationships. 
It is necessary to distinguish between dynamical laws and causality.
A unique feature of DLM is that supervenient causes arise in one dynamical system and effects occur in another dynamical system, and both dynamical systems share the same physical entities. 
Therefore, these two dynamical systems cannot be separated. When viewed as a single, integrated system, the cause originates within the system itself and affects it.
The key difference from general causal relationships is that these shared physical entities contribute to the asymmetry of causality. This unique feature can be characterized as ``intrinsic.'' 
It is a system-level causality that does not exist when broken down into its constituent elements.

\subsection{Clarification on supervenient causes and intervention}
It is crucial to note that, in the Dual-Laws Model, supervenient causes are not identified with the supervenience-subvenience relations themselves, nor with the supervenient functions as such. 
While each supervenient function is defined by a fixed supervenience–subvenience relation, supervenient causes are defined instead as changes in the equations constructed by selecting and ordering subsets of these supervenient functions.
Importantly, the selection and ordering of supervenient functions are not determined by the underlying supervenience–subvenience relations. 
This independence is essential for treating supervenient causes as genuine causes in the interventionist sense. 
At the supervenience level, it is possible to intervene on the construction of these equations—by altering which supervenient functions are selected or how they are combined—while holding the subvenient dynamics and supervenience relations fixed.
Such interventions systematically change the subsequent evolution of subvenient states via the self-referential feedback control mechanism. Therefore, although supervenient functions themselves supervene on subvenient entities, supervenient causes, defined as changes in equation construction, play an autonomous causal role at the higher level. 
This separation allows the model to represent hierarchical causation without collapsing supervenient causes into mere redescriptions of subvenient dynamics.

\subsection{The DLM and Anomalous Monism}
We define the correspondence between the DLM framework and the reconstructed Anomalous Monism, and examine whether DLM satisfies the conditions of Anomalous Monism.

In DLM, we define dynamics at the supervenience level as C-Mental Laws and dynamics at the subvenience level as C-Physical Laws. In this case, C-Physical Laws are influenced by supervenient causes. If we interpret changes at the supervenience level as C-Mental Events, then C-Physical Events are influenced by C-Mental Events. Therefore, this satisfies condition (1) of Anomalous Monism.

The laws governing C-Physical Events are not solely determined by C-Physical Events. Furthermore, since the C-Mental Event is governed by dynamical laws with inter-individual differences, it is not a strict law. Consequently, C-Physical Events influenced by supervenient causes also follow ``anomalous'' laws. Therefore, condition (3) is satisfied.

Finally, we examine condition (2). This condition alone establishes consistency between the Causal Stance and the Physical Stance. From the Causal Stance perspective, the Dual-Laws Model is distinguished from a single-layer dynamical system because asymmetry is distinguished. Conversely, when viewed from the Physical Stance, the asymmetry between cause and effect is ignored. 
In this case, supervenient causes are changes in equations composed of supervenient functions, all of which can be described as functions of subvenient entities. In other words, the changes of subvenient states are solely determined by the state changes of subvenient entities. Therefore, it satisfies P-Physical Determinism. 

When mental causation is modeled as inter-level causation realized by self-referential feedback mechanisms, it satisfies P-Physical Laws. In summary, under the Physical Stance, the DLM holds that changes in P-Physical States can be described solely in terms of P-Physical States and do not require nonphysical forces or energy. 
Furthermore, since supervenience level causally constrains the corresponding subvenient states, they embody the ''intrinsic'' or self-determining nature of mental causation (\cite{robb2023, ryan2020, shogren2015}). According to Kim (\cite{kim1998}), the supervenient-subvenient relation is the most promising way to characterize the mind from a physicalist perspective.
We assume that DLM satisfies condition (2) of Anomalous Monism.

It is not easy to define intrinsic causes. As with volition (\cite{haggard2018}), which is explained by intrinsic causes, it is easier to characterize it through negative statements. Volition excludes not only action caused by external stimuli but also those driven by physiological needs within the system. It is not enough to simply assume that intrinsic causes lie within the system itself. To explain intrinsic causation, inter-level causation and whole-to-parts causation have often been the focus of attention (\cite{sperry1991, kim1998}). We assume that whole-to-parts causation explains intrinsic causes because this is a system-level causal relationship that does not exist at the element level and is independent of extrinsic causes.

What is important in this discussion is that, in the Physical Stance, both C-Mental Laws and C-Physical Laws are interpreted as P-Physical Laws. In the Causal Stance, however, because it distinguishes the asymmetry in causality, it becomes possible to distinguish between C-Mental Laws and C-Physical Laws. In other words, C-Physical Laws are distinct from P-Physical Laws.
P-Physical States obey P-Physical Laws, and their evolution is determined solely by P-Physical States.
On the other hand, C-Physical Laws are the laws that C-Physical Events follow, and they are influenced by C-Mental Events.
Using a model that ignores the asymmetry in the Causal Stance, we can represent P-Physical Laws that satisfy physical determinism.

At present, there is no common language within the scientific community studying consciousness and the mind. This field lies at the boundary between mind and physics, and there is a possibility that the same word may have different meanings depending on the stance. As scientific fields mature, such confusion does not arise, so a notation method that explicitly specifies the stance may become unnecessary in the future. However, at this point in time, there is a high risk that meanings will become ambiguous and unclear unless the difference between the Physical Stance and the Causal Stance is made explicit.

\section{What is the mind in a linguistic framework?}
While reviewing the arguments presented in this paper, we will discuss methods by which scientific theories can incorporate practical mental terminology into their linguistic frameworks in accordance with their objectives.
Adopting a materialist stance and upholding physical determinism means that, if we wish to distinguish between the physical and the mental, as in folk psychology, we will have no choice but to redefine the two using a linguistic framework that is different from that of physics.
We believe that such a redefinition is necessary because there is a gap between the concept of ``physical'' in folk psychology and in physics.
Similar remarks can also be found in \cite{searl}. 

The term ``physical causal closure'' appears to be frequently used in philosophy (\cite{gebharter,kallestrup, kleiner, woodward2015, stern}). The major problem is that while Kim equates causation with physical determinism (\cite{kim1998}), the currently dominant definition of causation does not take this position. 
The term ``physical causal closure'' is a concept that is not definable within the Physical Stance alone unless one accepts Kim’s premises regarding causation. 
This is because physics does not define causation as an asymmetric, intervention-based relation.

In his book, Kim criticizes Anomalous Monism for saying little about the relationship between mind and physics. This stems from Davidson’s assumption that there are no strict laws governing the relationship between supervenience and subvenience. Therefore, our strategy differs from Davidson’s. We treat the relationship between supervenience and subvenience as the fixed merelogical relationship. If supervenience-level mental laws do not follow strict laws, then neither do subvenience-level physical laws. We resolve the resulting contradiction by distinguishing between the Physical Stance and the Causal Stance. We interpret P-Physical Laws as following strict laws, whereas C-Physical Laws do not.

We intuitively believe that we can act on our own volition. Davidson was strongly motivated to defend such action. There are several practical academic fields, such as the neuroscience of volition (\cite{haggard2018}), and self-determining theory in psychology (\cite{ryan2020}), that acknowledge mental causation. These fields require a language in which physical causal closure does not hold because they acknowledge the existence of mental causation.

The concept of physical causal closure is not clearly defined within a given academic discipline's language and should not be considered a fundamental principle. This is because physics itself does not define causality or require physical causal closure. The question is how to develop a language for disciplines that acknowledge mental causation while maintaining consistency with physical determinism. Assuming physical causal closure would prevent the development of practical languages for specific fields. For this reason, we have focused on Davidson’s Anomalous Monism because it is a form of materialism that acknowledges mental causation.

In the Causal Stance, we have posited two conditions that mental causation must satisfy: 1) C-Mental supervene on C-Physical; and 2) it is possible to introduce an asymmetry of cause and effect from the supervenience level to the subvenience level. 
The reason for considering a hierarchy composed of multiple supervenience-subvenience relations is that it is believed that causality cannot be defined within a single supervenience-subvenience relation (\cite{kim1998}).
It is important for defining mental causation that causation from the supervenience level to the subvenience level can be interpreted as ``intrinsic.''
We consider these to be minimal conditions, but there may be other ways to satisfy them besides the proposed DLM.

We believe that constructing a language capable of describing the causality of the mind, as required by the academic field, is an important task. The DLM model can provide a physical model of Anomalous Monism. Based on this, we believe that Anomalous Monism can say sufficiently much about the mind and the physical world. On the other hand, Davidson’s Anomalous Monism does not presuppose a hierarchical model. In that sense, the Dual-Laws Model is not only interpretable as Anomalous Monism but also provides a framework for describing the relationship between mind and physics in the Causal Stance. Furthermore, the DLM can serve as a practical model for academic fields that require an account of mental causation.

By modeling the supervenience level as a layer of mental processes and the subvenience level as a layer of C-Physical processes, we can model inter-level causality as mental causation. It is important to note that while the supervenience level includes supervenient functions that share physical entities with subvenient entities, it also includes other physical entities that are used to select and order the supervenient functions. When modeling the relationship between mind and matter, we consider it inevitable that additional physical entities in the supervenience level are involved in mental processes. 
We propose distinguishing between C-Mental and C-Physical processes based on asymmetry in the physical model; we are not introducing non-physical entities into mental processes.

In the Physical Stance, the mind cannot be defined, nor can we question what the mind is. 
However, if we consider the mind to be an entity independent of the physical world, then we cannot conceive of any interaction between the two. 
If the mind interacts with the physical world, then the mind must be physical. 
Yet, by definition, the mind is non-physical. 
To resolve this contradiction, we must consider two frameworks: the monistic Physical Stance and the dualistic Causal Stance. 
Here, the ``monistic'' and ``dualistic'' we refer to are descriptive, not ontological.
Our goal is to construct a linguistic framework for a field of study that deals with the mind and mental causation without contradicting physics; it is not to clarify the ontology of the mind.
To achieve this, we must also ensure that models within the Causal Stance can be translated into the Physical Stance.
Within the Physical Stance, the C-Mind can be translated into the P-Physical, which is capable of interacting with the physical world. 
On the other hand, the C-Mind is not C-Physical because, within the Causal Stance, the two are distinguished by causal asymmetry.

We believe that the question ``What exactly is the mind?'' is defined within the linguistic framework of a given academic discipline. If, as in physics, a common model of the mind could be developed that is highly universal and can be used as a standard across many applied sciences, that model would likely be the one that best reflects the actual nature of the mind. Our position is that the necessary conditions for modeling the mind are that it supervenes on the physical in the Causal Stance, that it can explain ``intrinsic'' mental causation in the Causal Stance, and that it does not contradict physical determinism in the Physical Stance. Furthermore, we believe this study has demonstrated that distinguishing between the Physical Stance and the Causal Stance is crucial for achieving this.

\section{Conclusion}
This study began with the premise that scientific communities possess distinct linguistic systems depending on their objectives. Next, focusing on the differences between the language of physics and that of causality, we proposed a distinction between Physical Stance and Causal Stance. Because Physical Stance ignores asymmetries based on manipulability, models that can be distinguished under Causal Stance may not be distinguishable under Physical Stance.

We introduced the distinction between the Physical Stance and the Causal Stance into the debate on mental causation in philosophy of mind. 
We then argued that Kim’s physical causal closure is undefined within the Physical Stance. 
We then proposed that it is possible to construct a causal model within the Causal Stance that satisfies the three conditions of Anomalous Monism. 
This implies that, within the Causal Stance, it is possible to construct models in which C-Physical Causal Closure is false. 
However, in such cases, it is necessary to distinguish clearly between P-Physical Laws and C-Physical Laws. 
In Davidson’s argument, no contradiction arises if the laws governing C-Physical Events, as defined within the Causal Stance, are interpreted as C-Physical Laws. 
However, as there is no evidence that he was aware of a clear distinction between physics and causality, 
We believe there is merit in clarifying the distinction between stances.

In many applied sciences, numerous disciplines require causality as a prerequisite, and Pearl’s framework for causal inference is used in these areas. But should the language of the still-developing science of the mind and consciousness include causality? Or should we only use the language of physics to explain it? 
If we consider the cause of consciousness to be a scientific question, we should adopt the Causal Stance. However, we cannot rule out the possibility that consciousness should be elucidated by adopting the Physical Stance, as with other physical phenomena. 
This study aims to clarify the distinction between the Physical Stance and the Causal Stance, demonstrating that P-Physical States are determined by other P-Physical States and that it is possible to construct a language in which C-Physical Causal Closure does not hold. 
We will leave it to future discussions to determine what stance the science of consciousness and the mind should adopt going forward.

\section*{Appendix}

We describe the formulation of DLM. 
The formulation assume that multiple supervenience-subvenience relationships. The supervenience level consists of multiple supervenient functions and the subvenience level consists of the corresponding subvenient entities.

They define subvenience level states as $SUB_i \subset \mathbb{R}^{n_i}$ for an index set $i \in I, I \subset \mathbb{N}$. They define the corresponding supervenient function whose domain is direct product of $N \subset \mathbb{N}$ sets $$\underbrace{ \mathbb{R}^m \times \mathbb{R}^m \times \dots \times \mathbb{R}^m}_N$$ and codomain is $\mathbb{R}^m$ as follows: $SUP[N]: \mathbb{R}^m \times \mathbb{R}^m \times \dots \times \mathbb{R}^m \to \mathbb{R}^m$. They define the bridge function $B: SUB \to SUP[N]$. Each supervenience-subveneince relation can be represented by $X^i = b_i (x^i), X^i \in SUP[N_i], x^i \in SUB, b_i \in B, i \in I$. 

Next, they define equations by selecting and ordering multiple supervenient functions from a set of supervenient functions $SUP = \bigcup_i SUP[N_i]$. 
Consider an index sequence $c = [i_0, i_1, ...] \in C$, where $i_k \in I$. 
Each index corresponds one-to-one with an element $SUP$. Let error function $E: \mathbb{R}^m \to \mathbb{R}^m$ and $S: C \to E$ be a mapping from an index sequence to error function. 
The error function can be represented by an algebraic formula of the supervenient functions, and the order of operations is determined by the index sequence. 
They assume that the equation is satisfied when the output of this error function is zero.
Array of multiple index sequences $\mathbf{c} = [c_0, c_1, ...]$ is a discrete supervenience-level state. 

They introduce causal relationship between the error function and index sequences. Using \cite{pearl}'s causal operator ``$:=$'' (an assignment operator), they proposed that the error function $e \in E$ can be described by $e:=s(c), s \in S$. 
The symbol $:=$ means ``is determined by,'' and Pearl himself uses $=$ to express this meaning.
Similarly, they defined multiple error functions for the array of index sequences $\mathbf{c}$. 
They argued that considering changes in index sequences as causes at the supervenience level is crucial to defining inter-level causation. Therefore, they proposed a model that incorporates causal semantics into determining the error function from the index sequence.
Without identifying the asymmetry between cause and effect, this operator cannot be introduced. 

Feedback error $err$ can be calculated by error function $e \in E$ and input $x \in \mathbb{R}^m$ like $err = e(x)$. 
The dynamics of subvenient states can be described as follows: $ x^i_{t+1}, err_{t+1} = p (x^i_t, err_t)$. 
The dynamics of index sequences can be described as follows: $\mathbf{c}_{T+1} = P (\mathbf{c}_{T})$. 

In this paper, we propose that the dynamical laws $p$ are C-Physical Laws and the dynamical laws $P$ are C-Mental Laws in Causal Stance. 
Here, the dynamical laws do not imply causal relationship. In this model, causal relationships are described by the assignment operator. 
In Physical Stance, the causal assignment operator $:=$ is replaced by $=$, because the semantics of physics does not distinguish the asymmetry between cause and effect. 
This model of Physical Stance follows physical determinism, as all subvenient states are determined solely by subvenient states, and thus requires no non-physical forces. In Physical Stance, this model can be described by $x_{T+1}= f_{c_t}(x_{t})$. 
In Causal Stance, the causal assignment operator serves as a distinguishing feature between the DLM and the single-level dynamical system model.

\subsection* {Acknowledgments}
This research was supported by the ******. The funding sources had no role in the decision to publish or prepare the manuscript.


\printbibliography

\end{spacing}
\end{document}